\documentclass[aps,twocolumn,preprintnumbers,floatfix,nofootinbib]{revtex4}

\usepackage{graphicx}% Include figure files
\usepackage{amsmath}
\usepackage{amsfonts}
\usepackage{amssymb}
\usepackage[utf8]{inputenc}
\usepackage{enumitem}
\usepackage{amsmath}
\usepackage{amsfonts}
\usepackage{amssymb}
\usepackage{color}
\usepackage{CJKutf8}

\def\be{\begin{equation}}
\def\ee{\end{equation}}
\def\bea{\begin{eqnarray}}
\def\eea{\end{eqnarray}}
\def\J{\mathcal{J}}

\def\tr{\mathrm{tr}}
\def\ordr{\mathcal{O}}

\def\<{\langle}
\def\>{\rangle}

\def\nlsm{nl$\sigma$m}

\def\N{$N$}
\def\non{\nonumber}

\def\alpd{{\dot{\alpha}}}

\def\tr{\mathrm{tr}}
\def\ordr{\mathcal{O}}

\def\ss{{\text{ss}}}
\def\nl{{\text{nl}}}

\def\slashchar#1{\setbox0=\hbox{$#1$}           % set a box for #1
   \dimen0=\wd0                                 % and get its size
   \setbox1=\hbox{/} \dimen1=\wd1               % get size of /
   \ifdim\dimen0>\dimen1                        % #1 is bigger
      \rlap{\hbox to \dimen0{\hfil/\hfil}}      % so center / in box
      #1                                        % and print #1
   \else                                        % / is bigger
      \rlap{\hbox to \dimen1{\hfil$#1$\hfil}}   % so center #1
      /                                         % and print /
   \fi}

\begin{document}

%\preprint{HUTP-05/A???}

\title{New Flavor-Kinematics Dualities and Extensions of Nonlinear Sigma Models}
%\vspace*{2cm}

\author{Ian Low$^{\, a,b}$ and Zhewei Yin$^{\, b}$}
%\email[e-mail: ]{}
\affiliation{%\vspace{0.5cm}
\mbox{$^a$ High Energy Physics Division, Argonne National Laboratory, Argonne, IL 60439, USA}\\
\mbox{$^b$ Department of Physics and Astronomy, Northwestern University, Evanston, IL 60208, USA} 
%\mbox{$^c$ Theoretical Physics Department, CERN, 1211 Geneva 23, Switzerland}
% \vspace{0.5cm}
%\vspace{1cm}
}

\begin{abstract}
%\vspace*{0.5cm}
Nonlinear sigma  model (\nlsm) based on the coset SU(\N)$\times$SU(\N)/SU(\N)  exhibits several intriguing features at the leading ${\cal O}(p^2)$ in the derivative expansion, such as the flavor-kinematics duality and an extended theory controlling the single and triple soft limits. In both cases the cubic biadjoint scalar  theory plays a prominent role. We extend these features in two directions. First we uncover a new extended theory for  SO($N+1$)/SO(\N) \nlsm\ at ${\cal O}(p^2)$, which is a  cubic bifundamental/biadjoint   scalar theory.  
Next we provide evidence for flavor-kinematics dualities up to ${\cal O}(p^4)$ for both SU(\N)  and SO(\N)  \nlsm 's. In particular, we introduce a new duality building block based on the symmetric tensor $\delta^{ab}$ and demonstrate several flavor-kinematics dualities for 4-point amplitudes, which precisely match the soft blocks employed to soft-bootstrap the \nlsm's up to $\ordr (p^4)$.

\end{abstract}

%\pacs{}

\maketitle

%%%%%%%%%%%%%%%%%%%%%%%%%%%%%%%%%%%%%
\section{Introduction}
\label{sec:introduction}
%%%%%%%%%%%%%%%%%%%%%%%%%%%%%%%%%%%%%

The \nlsm\ \cite{GellMann:1960np} is the prototype of effective field theories, with the general construction based on an arbitrary coset $G/H$ given in Refs.~\cite{Coleman:1969sm,Callan:1969sn}  exactly half-a-century ago.  
However, it was  relatively recently when several new and surprising aspects were uncovered. They include:

\noindent
%\item 
(\begin{CJK*}{UTF8}{bkai}一\end{CJK*}) An infrared (IR) construction without reference to the broken group $G$, implemented either at the Lagrangian level through shift symmetries \cite{Low:2014nga,Low:2014oga} or at the on-shell level through the soft recursion relation \cite{Cheung:2015ota,Cheung:2016drk,Elvang:2018dco}. The IR universality may have important implications for models attempting to understand the origin of electroweak symmetry breaking \cite{Liu:2018vel,Liu:2018qtb}. 

\noindent
%\item 
(\begin{CJK*}{UTF8}{bkai}二\end{CJK*}) For SU(\N) \nlsm\ there exists a larger, extended theory of biadjoint cubic scalars interacting with the Nambu-Goldstone bosons (NGB), which controls both the  single soft \cite{Cachazo:2016njl,Low:2017mlh} and triple soft \cite{Low:2018acv} limits at leading order in the derivative expansion.

\noindent
%\item 
(\begin{CJK*}{UTF8}{bkai}三\end{CJK*}) Again for SU(\N) \nlsm\ at ${\cal O}(p^2)$, tree amplitudes exhibit flavor-kinematics duality \cite{Bern:2008qj, Chen:2013fya,Chen:2014dfa,Chen:2016zwe,Cheung:2016prv,Cheung:2017yef,Mizera:2018jbh,Bern:2019prr} that is most transparent in the Cachazo-He-Yuan (CHY)  \cite{Cachazo:2013gna,Cachazo:2013hca,Cachazo:2013iea,Cachazo:2014xea} or the positive geometry  \cite{Arkani-Hamed:2017mur} representations of  \nlsm\ amplitudes. The cubic biadjoint scalar theory also plays a prominent role in these constructions.

%\end{itemize}
More broadly, these new developments belong to recent advances to understand quantum field theories from the IR \cite{ArkaniHamed:2008gz,Cheung:2014dqa,Cachazo:2015ksa,Du:2015esa,Du:2016njc,Arkani-Hamed:2016rak,Rodina:2016jyz,Strominger:2017zoo,Zlotnikov:2017ahq,Rodina:2018pcb,Yin:2018hht,Carrasco:2019qwr,Kampf:2019mcd} and/or on-shell perspectives \cite{Elvang:2013cua,Cheung:2017ems,Du:2018khm,Zhou:2018wvn,Bollmann:2018edb,Bjerrum-Bohr:2018jqe,Gomez:2019cik,Zhou:2019mbe}. Origins of many of the recently discovered features are still poorly understood, even for gauge theories and gravity, although certain insights can be obtained by supersymmetrization or taking the low-energy limit of string theories.  In this regard,  scalar theories such as the \nlsm\ may serve as a simpler playing field than gauge theories and gravity.

In this work we will focus on the flavor-kinematics duality and the extended theory  mentioned above. Our goal is two-fold: to understand whether these features persist outside the SU(\N)$\times$SU(\N)/SU(\N) coset and/or beyond the leading ${\cal O}(p^2)$ in the derivative expansion. We study both the SU(\N) and the SO(\N) \nlsm's. The latter, based on the SO($N+1$)/SO(\N) coset, is a new example for ``soft-bootstrap" introduced in Ref.~\cite{Low:2019ynd},  where the NGB transforms as the fundamental  of SO(\N). In Yang-Mills theories the gauge field always transforms as the adjoint representation. The SO(\N) \nlsm, which contains particles {\em only} in the fundamental representation, is a new venue for exploration.

Two previous works on flavor-kinematics duality applied the ${\cal O}(p^2)$ Bern-Carrasco-Johansson (BCJ) relations  \cite{Bern:2008qj}  to SU(\N) \nlsm\ amplitudes with higher derivatives and concluded  that the duality fails already at ${\cal O}(p^4)$ \cite{Elvang:2018dco,Carrillo-Gonzalez:2019aao}. It is further claimed that at ${\cal O}(p^6)$ there is a single operator satisfying the ${\cal O}(p^2)$ BCJ relation, which corresponds to a low-energy limit of open strings called the abelian $Z$ theory \cite{Carrasco:2016ldy,Carrasco:2016ygv}.  Recently Ref.~\cite{Carrasco:2019yyn} explored a more general construction of color-kinematics duality and applied to a special class of higher derivative operators in gravity and gauge theories. Does color/flavor-kinematics duality hold for generic higher dimensional operators from an effective field theory point of view? We will present evidence for a positive proposition, at least for \nlsm's at $\ordr (p^4)$.

In this  note we will present several results and sketch their derivations, which include the existence of a new extended theory for SO(\N) \nlsm\ involving cubic interactions of  bifundamental/biadjoint scalars, as well as evidence for new flavor-kinematics dualities up to $\ordr (p^4)$ for both SU(\N) and SO(\N) \nlsm's. Many details are saved for a future publication.

%%%%%%%%%%%%%%%%%%%%%%%%%%%%%%%%%%%%%
\section{All about That basis}
\label{sec:basis}
%%%%%%%%%%%%%%%%%%%%%%%%%%%%%%%%%%%%%

In this section we discuss the three flavor bases that will be used in this work: the single trace, the Del Duca-Dixon-Maltoni (DDM) \cite{DelDuca:1999rs} and the pair basis \cite{Low:2019ynd}. While the first two bases are commonly employed for amplitudes in Yang-Mills theories, we  adapt them to the general $G/H$ \nlsm. The pair basis, on the other hand, is specific to theories containing the fundamental of SO(\N).

For a general coset $G/H$ the unbroken generators $T^i$ in $H$ and the broken generators $X^a$ in $G/H$  satisfy
\be
\begin{split}
&[T^i, T^j]=if^{ijk} T^k \quad , [T^i, X^a] = if^{iab} X^b   \ ,\\
&  [X^a, X^b]=if^{abi}T^i +i f^{abc}X^c \ .
\end{split}
\ee
In a symmetric coset $f^{abc}=0$, which we will focus on.  For internal symmetry there is an NGB $\pi^a$ for each broken generator $X^a$. Therefore, the NGB carries the adjoint index of the broken group $G$. From this perspective, tree amplitudes of NGBs at the leading order can then be written naturally in the ``single trace" basis \cite{Kampf:2012fn, Kampf:2013vha,Bijnens:2019eze}:
\bea
&&\!\!\! M_n^{a_1 \cdots a_{n} } (p_1, \cdots, p_{n}) \equiv  \nonumber\\
&&\ \ \sum_{\alpha \in S_{n-1}} {\rm Tr}(X^{ a_{\alpha (1)}} \cdots X^{a_{\alpha (n-1)}} X^{a_n}) M_n( \alpha, n )\ ,\label{eq:stfod}
\eea
where $\alpha$ is a permutation of $\{1, \cdots, n-1\}$ and $M(\alpha,n)$ is the flavor-ordered partial amplitudes. Because of the cyclic invariance of the trace, there are $(n-1)!$ independent trace structures. Nevertheless, the partial amplitudes satisfy the Kleiss-Kuijf (KK) \cite{Kleiss:1988ne} and BCJ relations \cite{Bern:2008qj,Chen:2013fya}, which reduce the independent partial amplitudes to $(n-2)!$ and $(n-3)!$, respectively. Using the KK relations, the full amplitude can be written in the more minimal DDM basis, which for \nlsm\ amplitudes is
\bea
&&M_n^{a_1 \cdots a_n} =  \sum_{\alpha \in S_{n-2}} (-1)^{n/2-1} f^{a_1 a_{\alpha (1)} i_1 } \nonumber \\
 &&\times  \left( \prod_{j=1}^{n/2-2} f^{i_{j} a_{ \alpha (2j)} b_j} f^{b_j a_{\alpha (2j+1)} i_{j+1}} \right)  f^{i_{n/2-1} a_{\alpha (n-2)} a_n} \nonumber\\
&&\times \  M_n (1,\alpha, n)\ ,\label{eq:ddm}
\eea
where $\alpha $ is a permutation of $\{2,3,\cdots, n-1\}$. The DDM basis contains only $(n-2)!$ independent flavor structures.

For the SO($N+1$)/SO(\N) \nlsm, the SO($N+1$) structure constants $f^{iab}$ satisfy   the   relation
\bea
 f^{iab } f^{icd} =- \frac{1}{2}  \left(\delta^{ad} \delta^{bc} - \delta^{ac} \delta^{bd}\right)\ ,\label{eq:comrel}
\eea
which is secretly the completeness relation for generators in the fundamental representation of the SO(\N) group. The relation can be used to further simplify the DDM basis:
\bea
&&\!\!\!\!\!\!\!M_n^{a_1 \cdots a_n}(p_1, \cdots, p_{n}) =  \non\\
&&\!\!\!\!\!\!\!\!\!\!\!\!\!\!\sum_{\alpd \in P_n} \left( \prod_{j=1}^{n/2} \delta^{a_{\alpd_{2j-1}} a_{\alpd_{2j}}} \right) M_n (\alpd_1\alpd_2|\alpd_3\alpd_4| \cdots | \alpd_{n-1} \alpd_n),\label{eq:fdpair}
\eea
where $P_n$ is the  partition of $\{12 \cdots  n\}$ into $n/2$ subsets,  $\{\alpd_1\alpd_2\}, \{\alpd_3 \alpd_4\}, \cdots \{ \alpd_{n-1} \alpd_n\}$, where each subset contains two elements contracted by Kronecker delta $\delta^{\alpd_{2j-1} \alpd_{2j}}$ and therefore symmetric in $\alpd_{2j-1} \leftrightarrow \alpd_{2j}$. The argument of the partial amplitude $M_n (\alpd) $  then contains $n/2$ non-ordered pairs of external particle indices. Since we are considering a symmetric coset, the amplitude contains an even number of external legs. The right-hand side (RHS) of the above  is a sum of $(n-1)!!$ independent flavor factors and forms what we call the ``pair basis.''  Since $\delta^{ab} = \tr \left( T^a T^b \right)$, the pair basis is a multi-trace basis. The amplitude in this basis is invariant under exchanging the positions of different traces, as well as exchanging the two particle labels in each trace. For $n\ge 8$ the pair basis becomes smaller than the BCJ basis. In addition, it is possible to express partial amplitudes in the pair basis as linear combinations of those in the single trace basis. This relation leads to a CHY representation for all tree amplitudes in the pair basis of SO(\N) \nlsm.

%%%%%%%%%%%%%%%%%%%%%%%%%%%%%%%%%%%%%
\section{New Extension of NLSM}
%%%%%%%%%%%%%%%%%%%%%%%%%%%%%%%%%%%%

For a set of scalars $\pi^a$ transforming under the fundamental representation of the SO(\N) group, the leading ${\cal O}(p^2)$ Lagrangian satisfying the Adler's zero condition and describing the SO($N+1$)/SO(\N) \nlsm\ is \cite{Low:2019ynd}
\bea
{\cal L}^{(2)} &=& \frac{1}{2} F_1 (R)\ \partial_\mu\pi^a \partial^\mu \pi^a \nonumber\\
&&\quad - \frac{1}{4f^2 R} \left[ F_1 (R) - 1 \right] (\pi^a \partial_\mu \pi^a)^2\ ,\label{eq:lnsonf}
\eea
 where $F_1(x)\equiv (1/x) \sin^2\sqrt{x}$, $R\equiv \pi^a\pi^a/(2f^2)$ and $f$ is the Goldstone decay constant. The flavor structure of NGB's in the Lagrangian is such that all flavor indices are contracted pair-wise by the Kronecker deltas, leading to the pair basis introduced in Sect.~\ref{sec:basis}. Eq.~(\ref{eq:lnsonf}) is invariant under the  nonlinear shift symmetry,
 \be
 \pi^a \to \pi^a +\varepsilon^a +\frac1{2f^2R}\left[F_2(R)-1\right]\left(\pi^2\varepsilon^a - \pi\cdot \varepsilon\ \pi^a\right)\ ,
 \ee
 where $F_2(x)\equiv \sqrt{x} \, \cot \sqrt{x}$. The conserved current corresponding to the shift symmetry is
 \bea
 \label{eq:current}
\!\!\!\!\!\!\!\!\!\!\!\!\!\!\!\mathcal{J}^a_\mu &=& \partial_\mu \pi^a\non\\
& &\!\!\! \!\!\!\!  + \sum_{k=1}^\infty \frac{(-4)^k}{(2n+1)! } \left[R^{k} \partial_\mu \pi^a - \frac{R^{k-1}}{2f^2} (\pi\cdot \partial_\mu \pi) \pi^a \right] \, ,
\eea
 and the corresponding Ward identity is
\be \label{eq:wardid}
i \partial^\mu \< \Omega| \J^a_\mu (x) \prod_{i=1}^n \pi^{a_i} (x_i) |\Omega \>\sim 0  \ ,
\ee
 where we have neglected Schwinger terms on the RHS since they do not contribute when all particles become on-shell.

 Following the same procedure as in Ref.~\cite{Low:2017mlh}, we perform the LSZ reduction and take the on-shell limit on Eq.~(\ref{eq:wardid}). Then the $\partial_\mu\pi^a$ term in $\mathcal{J}^a_\mu$ contributes a single-particle pole and gives rise to the $(n+1)$-point (pt) on-shell amplitude, which is related to the matrix elements of the higher-order operators in Eq.~(\ref{eq:current}). More explicitly,
 \bea
\label{eq:quantward}
 &&\!\!\! M_{n+1}^{a_1 \cdots a_{n+1}} (p_1, \cdots ,p_{n+1}) \nonumber\\
&&\quad = \sum_{k=1}^{\infty} \<0| \tilde{O}^a_k(p_{n+1}) | \pi^{a_1}(p_1) \cdots \pi^{a_n}(p_n)\> \, ,
\eea
where
\bea
\label{eq:Okdef}
&&\tilde{O}^a_k(p_{n+1}) =  \int d^4 x\, e^{-i p_{n+1} \cdot x}  \frac{-(-4)^k}{(2k+1)!}\non\\
&&\qquad \times\ \partial_\mu \left[R^{k} \partial_\mu \pi^a - \frac{R^{k-1}}{2f^2} (\pi\cdot \partial_\mu \pi) \pi^a \right] \ .
\eea
Notice that $\tilde{O}^a_k(p_{n+1})$ gives rise to a $(2k+1)$-pt vertex, which in the pair basis can be written as
\be
\label{eq:vertex}
V(q_1|q_2 q_3| \cdots | q_{2k} q_{2k+1}) =- \frac{ (-4)^k (k-1)!}{2f^{2k} (2k)!} p_{n+1}\cdot q_1\ ,
\ee
and when entering Eq. (\ref{eq:quantward}) at tree level, the $q_i$ in the above are sum of subsets of the external momenta $p_1, \cdots ,p_n$. 
In taking the single soft limit, $p_{n+1}\to \tau p_{n+1}$, $\tau \to 0$, Eq.~(\ref{eq:vertex}) starts at the linear order in $\tau$, which makes the Adler's zero manifest and leads to the  single soft theorem,
\bea
&& \!\!\!\!\!\!\!\!\!\!\!\! M_{n+1} (n+1,1|23|45|\cdots | n-1,n)\non\\
&&\!\!\!\!\!\!\!\!\!\!\!\!=-\frac{\tau}{2  }  \,  \sum_{k=2}^n s_{k,n+1} \sum_{\substack{j=2\\j \ne k}}^{n} M_n (1jj_p|\alpd^{(j)}\, ||\, \bar{\bar{1}}\,\bar{j}\,\bar{\bar{k}}) + \ordr (\tau^2) ,\label{eq:slsspb}
\eea
where $j_p \equiv j+(-1)^j$, $\alpd^{(j)}$ is the partition $\{23 |45| \cdots | n-1,n\}$ with the pair $\{j, j_p\}$ removed.
In the above $M_n (1jj_p|\alpd^{(j)}\, ||\, \bar{\bar{1}}\,\bar{j}\,\bar{\bar{k}})$ denotes the $n$-pt amplitude of an extended theory containing two bifundamental scalars,  charged under SO(\N)$\times \text{SO}(\overline{N})$, in particle-1 and particle-$k$ as well as a biadjoint scalar  in particle-$j$.  The ordering to the right of $||$ denotes the flavor structure under $\text{SO}(\overline{N})$, where two external bifundamental scalars $\phi^{a_r\bar{\bar{a}}_r}, r=1,k,$ and one biadjoint scalar $\Phi^{j\bar{j}}$ are contracted by $ ({T}^{\bar{j}})_{\bar{\bar{a}}_1\bar{\bar{a}}_k}$. The ordering to the left of  $||$ denotes the flavor structure under SO(\N) where $(1jj_p)$  are contracted by
$(T^j)_{a_1a_{j_p}}$, while the pair-wise indices in $\alpd^{(j)}$ are  contracted by Kronecker deltas. 
Eq.~(\ref{eq:slsspb}) is to be contrasted with the result in SU(\N) \nlsm\ \cite{Cachazo:2016njl},
\bea
\label{eq:chysun}
M_{n+1}(1\cdots n+1) = \tau \sum_{i=2}^{n-1} s_{n+1,i}\,M_{n}(1\cdots n || \bar{1}\bar{i}\bar{n})\, \non\\
+\ordr (\tau^2),
\eea
which is controlled by an extended theory with cubic biadjoint scalars charged under SU(\N)$\times\text{SU} (\overline{N})$.

In the SO(\N) extended theory, odd-pt vertices containing two $\phi$'s, one $\Phi$ and an even number of NGB's are given by
\be
\label{eq:cubicvertex}
 (T^i)_{ab} (T^{\bar{i}})_{\bar{\bar{a}}\bar{\bar{b}}}\ \phi^{a\bar{\bar{a}}} \, \phi^{b\bar{\bar{b}}}\,  \Phi^{i\bar{i}}\ \sum_{n=0}^\infty \frac{(-4)^{n} }{(2(n+1))! } \left(\frac{\pi^c\pi^c}{2f^2}\right)^{n}, \!
\ee
where $(T^{i})_{a b}$ and $(T^{\bar{i}})_{\bar{\bar{a}} \bar{\bar{b}}}$ are generators of SO(\N) and $\text{SO}(\overline{N})$, respectively. Eq.~(\ref{eq:cubicvertex}) leads to the amplitudes in the extended theory appearing in Eq.~(\ref{eq:slsspb}), if one further assumes that all even-pt vertices involved in these amplitudes are identical to those in the SO(\N) \nlsm, much like in the extension of the SU(\N) theory.

As an example, let us consider the 6-pt amplitude of the $\text{SO} (N)$ \nlsm\ in the pair basis:
\bea
&&M_6(61|23|45) =- \frac{1}{4} \left[s_{23} s_{16} \left( \frac{1}{P^2_{235}} + \frac{1}{P^2_{234}}\right) \right.\non\\
&&+ s_{23} s_{45} \left( \frac{1}{P^2_{236}} + \frac{1}{P^2_{123}}\right) + s_{45} s_{16} \left( \frac{1}{P^2_{245}} + \frac{1}{P^2_{345}}\right)\non\\
&&\left. \phantom{\left( \frac{1}{P^2_{236}} + \frac{1}{P^2_{123}}\right)}-s_{23} - s_{45} - s_{16} \right],\label{eq:6}
\eea
where we have suppressed the coupling constants. When we take $p_6$ to be soft,  Eq.~(\ref{eq:slsspb}) with $n=5$ tells us that
\bea
&&M_6(61|23|45)=-\frac{\tau}{2 } \non\\
&&\times \left\{ s_{26} \left[ M_5 (132|45||\bar{\bar{1}} \bar{3} \bar{\bar{2}}) +M_5 (145|23||\bar{\bar{1}} \bar{4} \bar{\bar{2}}) + M_5 (154|23||\bar{\bar{1}} \bar{5} \bar{\bar{2}}) \right]\right.\non\\
&&+ s_{36} \left[ M_5 (123|45||\bar{\bar{1}} \bar{2} \bar{\bar{3}}) +M_5 (145|23||\bar{\bar{1}} \bar{4} \bar{\bar{3}}) + M_5 (154|23||\bar{\bar{1}} \bar{5} \bar{\bar{3}}) \right]\non\\
&&+s_{46} \left[ M_5 (123|45||\bar{\bar{1}} \bar{2} \bar{\bar{4}}) +M_5 (132|45||\bar{\bar{1}} \bar{3} \bar{\bar{4}}) + M_5 (154|23||\bar{\bar{1}} \bar{5} \bar{\bar{4}}) \right]\non\\
&&\left.+ s_{56} \left[ M_5 (123|45||\bar{\bar{1}} \bar{2} \bar{\bar{5}}) +M_5 (132|45||\bar{\bar{1}} \bar{3} \bar{\bar{5}}) + M_5 (145|23||\bar{\bar{1}} \bar{4} \bar{\bar{5}}) \right] \right\}\non\\
&&+ \ordr (\tau^2),\label{eq:6ptst}
\eea
where $P_{ijk} = p_i + p_j + p_k$, and the amplitudes $M_5$ are of the following two different types:
\bea
M_5(123|45||\bar{\bar{1}} \bar{2} \bar{\bar{3}}) &=& -\frac{ s_{45}}{2 } \left( \frac{1}{s_{12}} + \frac{1}{s_{23}} \right),\label{eq:51}\\
M_5(123|45||\bar{\bar{1}} \bar{2} \bar{\bar{4}}) &=& -\frac{1}{2 }  \left( \frac{s_{45}}{s_{12}} - \frac{s_{25}}{s_{14}} \right).\label{eq:52}
\eea
Relabeling Eqs. (\ref{eq:51}) and (\ref{eq:52}) and plugging into Eq. (\ref{eq:6ptst}), we arrive at the result  consistent with Eq. (\ref{eq:6}).

%%%%%%%%%%%%%%%%%%%%%%%%%%%%%%%%%
\section{New Flavor-kinematics Dualities}
%%%%%%%%%%%%%%%%%%%%%%%%%%%%%%%%%

%%%%%%%%%%%%%%%%%%%%%%%%%%%%%%%%%
\begin{figure}[t]
\centering
\includegraphics[width=0.25\textwidth]{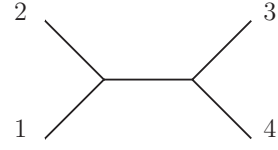}
\caption{The $s$-channel 4-pt cubic graph.\label{fig:4ptcub}}
\end{figure}
%%%%%%%%%%%%%%%%%%%%%%%%%%%%%%%%%
The flavor-kinematics duality hinges on, for the 4-pt cubic graph shown in Fig. \ref{fig:4ptcub}, a function of three flavor indices  $\textsf{j}(123)$ which satisfies anti-symmetry and the Jacobi identity \cite{Bern:2019prr}, 
\be
\textsf{j}(123) = -\textsf{j}(213)\ , \quad \textsf{j}(123) + \textsf{j}(231) + \textsf{j}(312) = 0\ .
\ee
Using Mandelstam variables we have  $(s,t,u)=(s_{12},s_{23},s_{13})$, where $s_{ij}=2p_i\cdot p_j$. Then $\textsf{j}_s \equiv \textsf{j}(123), \textsf{j}_t\equiv \textsf{j}(231)$ and $\textsf{j}_u \equiv \textsf{j}(312)$. In general $\textsf{j}(123)$ could be a function of both flavor factors and kinematic invariants \cite{Carrasco:2019yyn}.

Below we consider such functions that are local in momenta. At the lowest mass dimension, $\textsf{j}$ only contains flavor factors and the most well-known example is
\bea
\textsf{f}_A (123) = f^{ia_1 a_2} f^{ia_3 a_4} \ ,\label{eq:adjointc}
\eea
which involves the structure constant of a Lie group $H$ and is commonly employed for particles transforming under the adjoint representation such as Yang-Mills theories \cite{Bern:2019prr}. Eq.~(\ref{eq:adjointc}) is the special case of a more general situation. Consider  an arbitrary representation \textsf{r} of $H$ and choose a basis for the  generator $T_{ \textsf{r}}^i$ such that it is purely imaginary and anti-symmetric. If $(T^i_{ \textsf{r}})_{ab}$ satisfies the following ``Closure Condition'' \cite{Low:2014nga}
\be
(T_{ \textsf{r}}^i)_{a_1 a_2} (T^i_{ \textsf{r}})_{a_3 a_4}+ (T_{ \textsf{r}}^i)_{a_2 a_3} (T^i_{ \textsf{r}})_{a_1 a_4}+ (T_{ \textsf{r}}^i)_{a_3 a_1} (T^i_{ \textsf{r}})_{a_2 a_4}=0
\ee
then the flavor factor
\be
\label{eq:arbRc}
\textsf{f}_{\textsf{r}} (123) = (T_{\textsf{r}}^i)_{a_1 a_2} (T^i_{\textsf{r}})_{a_3 a_4}\ 
\ee
also satisfies anti-symmetry and Jacobi identity. When ${\textsf{r}}$ is the adjoint representation, Eq.~(\ref{eq:arbRc}) reduces to $\textsf{f}_A$. The Closure Condition is the condition that $\textsf{r}$ of $H$ can be embedded into a symmetric coset $G/H$ \cite{Low:2014nga}. For example,  the fundamental representation of SO(\N) is isomorphic to SO($N+1$)/SO(N) and the fundamental representation of SU(\N) parameterizes  SU($N+1$)/SU(\N)$\times$U(1).

At this mass dimension, however, there is a second possibility for $\textsf{j}(123)$  that has yet to be considered in the literature. It involves the rank-2 symmetric tensor $\delta^{ab}$, 
\bea
\textsf{f}_{\delta} (123) = \delta^{a_1 a_3} \delta^{a_2 a_4} - \delta^{a_1 a_4} \delta^{a_2 a_3}\ ,
\eea
which exists for {\em any} representations. For the fundamental representation of SO(\N), $\textsf{f}_{\delta} \propto \textsf{f}_{\textsf{r}}$ upon the completeness relation in Eq.~(\ref{eq:comrel}). For other groups/representations, it is a  new building block.

At the next order in  mass dimension, there is one building block which  contains only kinematic invariants,
\be
\textsf{n}^{\ss} (123) = t-u\ .
\ee
This turns out to be the kinematic numerator for the single-flavor Yang-Mills scalar theory.

These simple building blocks allow us to construct more complicated \textsf{j}'s.  One way is to just multiply existing blocks with objects invariant under permutations of (1234). For local building blocks, there are two such objects at the lowest mass dimension, without any kinematic invariants,
\bea
\textsf{d}_4^{abcd} &=&  \sum_{\sigma \in S_3} {\rm Tr} \left( T^{a_{\sigma (1)}} T^{a_{\sigma (2)}} T^{a_{\sigma (3)}} T^{a_4} \right),\\
\textsf{d}_2^{abcd} &=& \frac{1}{2} \sum_{\sigma \in S_3} \delta^{a_{\sigma (1)} a_{\sigma (2)}} \delta^{a_{\sigma (3)}a_4} .
\eea
Notice that $\textsf{d}_4$ is only for adjoint representations, while $\textsf{d}_2$ applies to any representation. For an arbitrary representation  $\textsf{d}_4$ can be generalized to the rank-$4$ totally symmetric tensor that is independent of $\textsf{d}_2$, if it exists. For example, the fundamental of SO(\N) does not have such a rank-4 symmetric tensor. Going to higher orders in mass dimension one can write down two permutation invariant building blocks,
\be
\textsf{X} = stu\ ,\  \quad \textsf{Y} =  s^2 + t^2 + u^2\ ,
\ee
which are first pointed out in Ref.~\cite{Carrasco:2019yyn}.

A second way to  generate new numerators is to take two existing blocks $\textsf{j}$ and $\textsf{j}^\prime$, and define $\textsf{J}_s (\textsf{j},\textsf{j}^\prime) = \textsf{j}_t \ \textsf{j}_t^\prime - \textsf{j}_u\ \textsf{j}_u^\prime$. Then $\textsf{j}''(1,2,3) = \textsf{J}_s (\textsf{j},\textsf{j}^\prime)$ is a  valid BCJ numerator \cite{Carrasco:2019yyn}. Using this technique one can build more numerators only containing momenta. For example,
\bea
\textsf{J}_s (\textsf{n}^\ss, \textsf{n}^\ss) \propto \frac{1}{3}  s(t-u)\equiv \textsf{n}^\nl (1,2,3)\ ,
\eea
which is the BCJ kinematic numerator for \nlsm\ at ${\cal O}(p^2)$ \cite{Du:2016tbc,Carrasco:2016ldy,Du:2017kpo}. 
However,  other  numerators constructed this way that only contain kinematic invariants  can all be written as a linear combination of $\textsf{n}^\ss$ and $\textsf{n}^\nl$ dressed with powers of  $\textsf{X}$ and $\textsf{Y}$ \cite{Carrasco:2019yyn}.

Focusing on 4-pt amplitudes in the remainder of this work, flavor-kinematics duality at the leading $\ordr (p^2)$ for both SU(\N) and SO(\N) \nlsm 's can be written as 
\bea
M^{(2)}_4 = \frac{1}{f^2} \left(\frac{\textsf{f}_{\textsf{r}\,s} \textsf{n}^\nl_s}{s} +\frac{\textsf{f}_{\textsf{r}\,t} \textsf{n}^\nl_t}{t} + \frac{\textsf{f}_{\textsf{r}\,u} \textsf{n}^\nl_u}{u} \right)\ ,\label{eq:nlfa2}
\eea
where $\textsf{r}= \text{Adjoint}$ for SU(\N) and Fundamental for SO(\N). For the SO(\N) \nlsm,  $\textsf{f}_{\textsf{r}}$ is also interchangeable with $\textsf{f}_{\delta}$. In both cases replacing $\textsf{f}_{\textsf{r}}$ by another copy of $\textsf{n}^\nl$ gives rise to the 4-pt amplitude, $M_4^{\text{sG}} \propto stu$, in special Galileon theory \cite{Cachazo:2014xea,Cheung:2014dqa}. Alternatively, one can realize  the double copy structure using the Kawai-Lewellen-Tye (KLT) relations \cite{Kawai:1985xq}.

At higher orders in derivative expansion, previous works aimed to generate {\em local} 4-pt amplitudes by modifying the kinematic factors $\textsf{n}^\nl$ while keeping the flavor factors $\textsf{f}_{\textsf{r}}$ intact, which turned out to be impossible until $\ordr (p^6)$ \cite{Elvang:2018dco,Carrillo-Gonzalez:2019aao}. Even at that order, the corresponding operator is a special one, belonging to the low-energy $\alpha^\prime$ expansion of open strings \cite{Carrasco:2016ldy}. In the following we will demonstrate a more general procedure to implement flavor-kinematics duality, which is valid for generic operators up to $\ordr (p^4)$.

It is convenient to recall that 4-pt amplitudes  at $\ordr (p^4)$ in \nlsm\ are characterized by four ``soft blocks" \cite{Low:2019ynd}, which are 4-pt local partial amplitudes in the single or double trace basis:
\begin{align}
&&  {\cal S}_1^{(4)} (1234) &=  \frac{u^2}{\Lambda^2 f^2}  \ ,   & {\cal S}_2^{(4)} (1234)&=   \frac{ st}{\Lambda^2 f^2} \ ,\label{eq:sa4s}\\
&& {\cal S}_1^{(4)} (12|34) &=   \frac{s^2}{\Lambda^2 f^2}   \ ,    & {\cal S}_2^{(4)} (12|34)&=    \frac{tu}{\Lambda^2 f^2} \ ,
\end{align}
where $\Lambda$ is a dimensionful parameter controlling the convergence of the derivative expansion. It turns out that there are a total of four possibilities to replace  $\textsf{f}_{\textsf{r}}$ that result in local 4-pt amplitudes at $\ordr (p^4)$, which correspond to the four soft blocks above. They are
\begin{align}
&& \hat{\textsf{f}}_1&= \frac{1}{\Lambda^2} \textsf{J} (\textsf{f}_{\textsf{r}},\textsf{n}^\ss)  \ ,   & \hat{\textsf{f}}_2&=\frac{1}{\Lambda^2} \textsf{d}_4^{a_1 a_2 a_3 a_4} \textsf{n}^\ss \ ,\label{eq:fh1234}\\
&& \hat{\textsf{f}}_3&= \frac{1}{\Lambda^2} \textsf{J} (\textsf{f}_{\delta},\textsf{n}^\ss)   \ ,    & \hat{\textsf{f}}_4&=\frac{1}{\Lambda^2} \textsf{d}_2^{a_1 a_2 a_3 a_4} \textsf{n}^\ss \ .
\end{align}
Replacing $\textsf{f}_{\textsf{r}}$ in Eq.~(\ref{eq:nlfa2}) by the above gives rise to four different $\ordr (p^4)$ amplitudes, whose partial amplitudes are related to the soft blocks via
\bea
M^{(4)}_{\hat{\textsf{f}}_1} (1234) &=& \ {\cal S}_1^{(4)} (1234) + 2\ {\cal S}_2^{(4)} (1234)\ , \\
M^{(4)}_{\hat{\textsf{f}}_2} (1234) &=& 2\left[ {\cal S}_1^{(4)} (1234)  - {\cal S}_2^{(4)} (1234) \right] \ ,\\
M^{(4)}_{\hat{\textsf{f}}_3} (12|34) &=& \ {\cal S}_1^{(4)} (12|34) + 2\ {\cal S}_2^{(4)} (12|34) \ ,\\
M^{(4)}_{\hat{\textsf{f}}_4} (12|34) &=& 2 \left[{\cal S}_1^{(4)} (12|34)  - {\cal S}_2^{(4)} (12|34) \right] \ .
\eea
Since the soft blocks generate tree amplitudes for the complete set of parity-even  operators at $\ordr (p^4)$, so does our construction of flavor-kinematics duality.

\section{Conclusion and Outlooks}
In this work we have demonstrated that several intriguing features persist beyond the leading $\ordr (p^2)$  effective action based on the SU(\N) \nlsm. We supplied a new example of extended theory in the SO(\N) \nlsm, also at the $\ordr (p^2)$, and presented evidence for flavor-kinematics duality at $\ordr (p^4)$ for both SU(\N) and SO(\N) \nlsm's, by providing a new duality building block based on $\delta^{ab}$, which could also potentially be applied to Yang-Mills theories.

Several future directions are currently being undertaken. They include amplitude relations and double-copy procedures in the pair basis, corrections to the single and double soft theorems from $\ordr (p^4)$ operators, the possible existence of an extended theory at $\ordr (p^4)$, and flavor-kinematics duality for $\ordr (p^4)$ amplitudes at higher multiplicities,    just to name a few. The outcome, as well as detailed derivations of results outlined in this note, will be reported in a future publication.
 
\begin{acknowledgments}
 We would like to thank John Joseph M. Carrasco and Laurentiu Rodina for collaborations on closely related subjects as well as useful comments on the manuscript.  Discussions with Nima Arkani-Hamed, Mariana Carrillo González, Callum R.T. Jones, Shruti Paranjape, Jaroslav Trnka, Mark Trodden and Suna Zekioglu are also acknowledged. This work is supported in part by the U.S. Department of Energy under contracts No. DE-AC02-06CH11357 and No. DE-SC0010143.  
\end{acknowledgments}

%%%%%%%%%%
%%%%%%%%%%    References
%%%%%%%%%%

%\bibliographystyle{utphys}
\bibliography{references_Op4_DC}

\begin{thebibliography}{60}
\expandafter\ifx\csname natexlab\endcsname\relax\def\natexlab#1{#1}\fi
\expandafter\ifx\csname bibnamefont\endcsname\relax
  \def\bibnamefont#1{#1}\fi
\expandafter\ifx\csname bibfnamefont\endcsname\relax
  \def\bibfnamefont#1{#1}\fi
\expandafter\ifx\csname citenamefont\endcsname\relax
  \def\citenamefont#1{#1}\fi
\expandafter\ifx\csname url\endcsname\relax
  \def\url#1{\texttt{#1}}\fi
\expandafter\ifx\csname urlprefix\endcsname\relax\def\urlprefix{URL }\fi
\providecommand{\bibinfo}[2]{#2}
\providecommand{\eprint}[2][]{\url{#2}}

\bibitem[{\citenamefont{Gell-Mann and Levy}(1960)}]{GellMann:1960np}
\bibinfo{author}{\bibfnamefont{M.}~\bibnamefont{Gell-Mann}} \bibnamefont{and}
  \bibinfo{author}{\bibfnamefont{M.}~\bibnamefont{Levy}},
  \bibinfo{journal}{Nuovo Cim.} \textbf{\bibinfo{volume}{16}},
  \bibinfo{pages}{705} (\bibinfo{year}{1960}).

\bibitem[{\citenamefont{Coleman et~al.}(1969)\citenamefont{Coleman, Wess, and
  Zumino}}]{Coleman:1969sm}
\bibinfo{author}{\bibfnamefont{S.~R.} \bibnamefont{Coleman}},
  \bibinfo{author}{\bibfnamefont{J.}~\bibnamefont{Wess}}, \bibnamefont{and}
  \bibinfo{author}{\bibfnamefont{B.}~\bibnamefont{Zumino}},
  \bibinfo{journal}{Phys. Rev.} \textbf{\bibinfo{volume}{177}},
  \bibinfo{pages}{2239} (\bibinfo{year}{1969}).

\bibitem[{\citenamefont{Callan et~al.}(1969)\citenamefont{Callan, Coleman,
  Wess, and Zumino}}]{Callan:1969sn}
\bibinfo{author}{\bibfnamefont{C.~G.} \bibnamefont{Callan},
  \bibfnamefont{Jr.}}, \bibinfo{author}{\bibfnamefont{S.~R.}
  \bibnamefont{Coleman}},
  \bibinfo{author}{\bibfnamefont{J.}~\bibnamefont{Wess}}, \bibnamefont{and}
  \bibinfo{author}{\bibfnamefont{B.}~\bibnamefont{Zumino}},
  \bibinfo{journal}{Phys. Rev.} \textbf{\bibinfo{volume}{177}},
  \bibinfo{pages}{2247} (\bibinfo{year}{1969}).

\bibitem[{\citenamefont{Low}(2015{\natexlab{a}})}]{Low:2014nga}
\bibinfo{author}{\bibfnamefont{I.}~\bibnamefont{Low}}, \bibinfo{journal}{Phys.
  Rev.} \textbf{\bibinfo{volume}{D91}}, \bibinfo{pages}{105017}
  (\bibinfo{year}{2015}{\natexlab{a}}), \eprint{1412.2145}.

\bibitem[{\citenamefont{Low}(2015{\natexlab{b}})}]{Low:2014oga}
\bibinfo{author}{\bibfnamefont{I.}~\bibnamefont{Low}}, \bibinfo{journal}{Phys.
  Rev.} \textbf{\bibinfo{volume}{D91}}, \bibinfo{pages}{116005}
  (\bibinfo{year}{2015}{\natexlab{b}}), \eprint{1412.2146}.

\bibitem[{\citenamefont{Cheung et~al.}(2016)\citenamefont{Cheung, Kampf,
  Novotny, Shen, and Trnka}}]{Cheung:2015ota}
\bibinfo{author}{\bibfnamefont{C.}~\bibnamefont{Cheung}},
  \bibinfo{author}{\bibfnamefont{K.}~\bibnamefont{Kampf}},
  \bibinfo{author}{\bibfnamefont{J.}~\bibnamefont{Novotny}},
  \bibinfo{author}{\bibfnamefont{C.-H.} \bibnamefont{Shen}}, \bibnamefont{and}
  \bibinfo{author}{\bibfnamefont{J.}~\bibnamefont{Trnka}},
  \bibinfo{journal}{Phys. Rev. Lett.} \textbf{\bibinfo{volume}{116}},
  \bibinfo{pages}{041601} (\bibinfo{year}{2016}), \eprint{1509.03309}.

\bibitem[{\citenamefont{Cheung et~al.}(2017)\citenamefont{Cheung, Kampf,
  Novotny, Shen, and Trnka}}]{Cheung:2016drk}
\bibinfo{author}{\bibfnamefont{C.}~\bibnamefont{Cheung}},
  \bibinfo{author}{\bibfnamefont{K.}~\bibnamefont{Kampf}},
  \bibinfo{author}{\bibfnamefont{J.}~\bibnamefont{Novotny}},
  \bibinfo{author}{\bibfnamefont{C.-H.} \bibnamefont{Shen}}, \bibnamefont{and}
  \bibinfo{author}{\bibfnamefont{J.}~\bibnamefont{Trnka}},
  \bibinfo{journal}{JHEP} \textbf{\bibinfo{volume}{02}}, \bibinfo{pages}{020}
  (\bibinfo{year}{2017}), \eprint{1611.03137}.

\bibitem[{\citenamefont{Elvang et~al.}(2019)\citenamefont{Elvang, Hadjiantonis,
  Jones, and Paranjape}}]{Elvang:2018dco}
\bibinfo{author}{\bibfnamefont{H.}~\bibnamefont{Elvang}},
  \bibinfo{author}{\bibfnamefont{M.}~\bibnamefont{Hadjiantonis}},
  \bibinfo{author}{\bibfnamefont{C.~R.~T.} \bibnamefont{Jones}},
  \bibnamefont{and}
  \bibinfo{author}{\bibfnamefont{S.}~\bibnamefont{Paranjape}},
  \bibinfo{journal}{JHEP} \textbf{\bibinfo{volume}{01}}, \bibinfo{pages}{195}
  (\bibinfo{year}{2019}), \eprint{1806.06079}.

\bibitem[{\citenamefont{Liu et~al.}(2018)\citenamefont{Liu, Low, and
  Yin}}]{Liu:2018vel}
\bibinfo{author}{\bibfnamefont{D.}~\bibnamefont{Liu}},
  \bibinfo{author}{\bibfnamefont{I.}~\bibnamefont{Low}}, \bibnamefont{and}
  \bibinfo{author}{\bibfnamefont{Z.}~\bibnamefont{Yin}},
  \bibinfo{journal}{Phys. Rev. Lett.} \textbf{\bibinfo{volume}{121}},
  \bibinfo{pages}{261802} (\bibinfo{year}{2018}), \eprint{1805.00489}.

\bibitem[{\citenamefont{Liu et~al.}(2019)\citenamefont{Liu, Low, and
  Yin}}]{Liu:2018qtb}
\bibinfo{author}{\bibfnamefont{D.}~\bibnamefont{Liu}},
  \bibinfo{author}{\bibfnamefont{I.}~\bibnamefont{Low}}, \bibnamefont{and}
  \bibinfo{author}{\bibfnamefont{Z.}~\bibnamefont{Yin}},
  \bibinfo{journal}{JHEP} \textbf{\bibinfo{volume}{05}}, \bibinfo{pages}{170}
  (\bibinfo{year}{2019}), \eprint{1809.09126}.

\bibitem[{\citenamefont{Cachazo et~al.}(2016)\citenamefont{Cachazo, Cha, and
  Mizera}}]{Cachazo:2016njl}
\bibinfo{author}{\bibfnamefont{F.}~\bibnamefont{Cachazo}},
  \bibinfo{author}{\bibfnamefont{P.}~\bibnamefont{Cha}}, \bibnamefont{and}
  \bibinfo{author}{\bibfnamefont{S.}~\bibnamefont{Mizera}},
  \bibinfo{journal}{JHEP} \textbf{\bibinfo{volume}{06}}, \bibinfo{pages}{170}
  (\bibinfo{year}{2016}), \eprint{1604.03893}.

\bibitem[{\citenamefont{Low and Yin}(2018{\natexlab{a}})}]{Low:2017mlh}
\bibinfo{author}{\bibfnamefont{I.}~\bibnamefont{Low}} \bibnamefont{and}
  \bibinfo{author}{\bibfnamefont{Z.}~\bibnamefont{Yin}},
  \bibinfo{journal}{Phys. Rev. Lett.} \textbf{\bibinfo{volume}{120}},
  \bibinfo{pages}{061601} (\bibinfo{year}{2018}{\natexlab{a}}),
  \eprint{1709.08639}.

\bibitem[{\citenamefont{Low and Yin}(2018{\natexlab{b}})}]{Low:2018acv}
\bibinfo{author}{\bibfnamefont{I.}~\bibnamefont{Low}} \bibnamefont{and}
  \bibinfo{author}{\bibfnamefont{Z.}~\bibnamefont{Yin}},
  \bibinfo{journal}{JHEP} \textbf{\bibinfo{volume}{10}}, \bibinfo{pages}{078}
  (\bibinfo{year}{2018}{\natexlab{b}}), \eprint{1804.08629}.

\bibitem[{\citenamefont{Bern et~al.}(2008)\citenamefont{Bern, Carrasco, and
  Johansson}}]{Bern:2008qj}
\bibinfo{author}{\bibfnamefont{Z.}~\bibnamefont{Bern}},
  \bibinfo{author}{\bibfnamefont{J.~J.~M.} \bibnamefont{Carrasco}},
  \bibnamefont{and}
  \bibinfo{author}{\bibfnamefont{H.}~\bibnamefont{Johansson}},
  \bibinfo{journal}{Phys. Rev.} \textbf{\bibinfo{volume}{D78}},
  \bibinfo{pages}{085011} (\bibinfo{year}{2008}), \eprint{0805.3993}.

\bibitem[{\citenamefont{Chen and Du}(2014)}]{Chen:2013fya}
\bibinfo{author}{\bibfnamefont{G.}~\bibnamefont{Chen}} \bibnamefont{and}
  \bibinfo{author}{\bibfnamefont{Y.-J.} \bibnamefont{Du}},
  \bibinfo{journal}{JHEP} \textbf{\bibinfo{volume}{01}}, \bibinfo{pages}{061}
  (\bibinfo{year}{2014}), \eprint{1311.1133}.

\bibitem[{\citenamefont{Chen et~al.}(2015)\citenamefont{Chen, Du, Li, and
  Liu}}]{Chen:2014dfa}
\bibinfo{author}{\bibfnamefont{G.}~\bibnamefont{Chen}},
  \bibinfo{author}{\bibfnamefont{Y.-J.} \bibnamefont{Du}},
  \bibinfo{author}{\bibfnamefont{S.}~\bibnamefont{Li}}, \bibnamefont{and}
  \bibinfo{author}{\bibfnamefont{H.}~\bibnamefont{Liu}},
  \bibinfo{journal}{JHEP} \textbf{\bibinfo{volume}{03}}, \bibinfo{pages}{156}
  (\bibinfo{year}{2015}), \eprint{1412.3722}.

\bibitem[{\citenamefont{Chen et~al.}(2016)\citenamefont{Chen, Li, and
  Liu}}]{Chen:2016zwe}
\bibinfo{author}{\bibfnamefont{G.}~\bibnamefont{Chen}},
  \bibinfo{author}{\bibfnamefont{S.}~\bibnamefont{Li}}, \bibnamefont{and}
  \bibinfo{author}{\bibfnamefont{H.}~\bibnamefont{Liu}} (\bibinfo{year}{2016}),
  \eprint{1609.01832}.

\bibitem[{\citenamefont{Cheung and Shen}(2017)}]{Cheung:2016prv}
\bibinfo{author}{\bibfnamefont{C.}~\bibnamefont{Cheung}} \bibnamefont{and}
  \bibinfo{author}{\bibfnamefont{C.-H.} \bibnamefont{Shen}},
  \bibinfo{journal}{Phys. Rev. Lett.} \textbf{\bibinfo{volume}{118}},
  \bibinfo{pages}{121601} (\bibinfo{year}{2017}), \eprint{1612.00868}.

\bibitem[{\citenamefont{Cheung et~al.}(2018{\natexlab{a}})\citenamefont{Cheung,
  Remmen, Shen, and Wen}}]{Cheung:2017yef}
\bibinfo{author}{\bibfnamefont{C.}~\bibnamefont{Cheung}},
  \bibinfo{author}{\bibfnamefont{G.~N.} \bibnamefont{Remmen}},
  \bibinfo{author}{\bibfnamefont{C.-H.} \bibnamefont{Shen}}, \bibnamefont{and}
  \bibinfo{author}{\bibfnamefont{C.}~\bibnamefont{Wen}},
  \bibinfo{journal}{JHEP} \textbf{\bibinfo{volume}{04}}, \bibinfo{pages}{129}
  (\bibinfo{year}{2018}{\natexlab{a}}), \eprint{1709.04932}.

\bibitem[{\citenamefont{Mizera and Skrzypek}(2018)}]{Mizera:2018jbh}
\bibinfo{author}{\bibfnamefont{S.}~\bibnamefont{Mizera}} \bibnamefont{and}
  \bibinfo{author}{\bibfnamefont{B.}~\bibnamefont{Skrzypek}},
  \bibinfo{journal}{JHEP} \textbf{\bibinfo{volume}{10}}, \bibinfo{pages}{018}
  (\bibinfo{year}{2018}), \eprint{1809.02096}.

\bibitem[{\citenamefont{Bern et~al.}(2019)\citenamefont{Bern, Carrasco,
  Chiodaroli, Johansson, and Roiban}}]{Bern:2019prr}
\bibinfo{author}{\bibfnamefont{Z.}~\bibnamefont{Bern}},
  \bibinfo{author}{\bibfnamefont{J.~J.} \bibnamefont{Carrasco}},
  \bibinfo{author}{\bibfnamefont{M.}~\bibnamefont{Chiodaroli}},
  \bibinfo{author}{\bibfnamefont{H.}~\bibnamefont{Johansson}},
  \bibnamefont{and} \bibinfo{author}{\bibfnamefont{R.}~\bibnamefont{Roiban}}
  (\bibinfo{year}{2019}), \eprint{1909.01358}.

\bibitem[{\citenamefont{Cachazo
  et~al.}(2014{\natexlab{a}})\citenamefont{Cachazo, He, and
  Yuan}}]{Cachazo:2013gna}
\bibinfo{author}{\bibfnamefont{F.}~\bibnamefont{Cachazo}},
  \bibinfo{author}{\bibfnamefont{S.}~\bibnamefont{He}}, \bibnamefont{and}
  \bibinfo{author}{\bibfnamefont{E.~Y.} \bibnamefont{Yuan}},
  \bibinfo{journal}{Phys. Rev.} \textbf{\bibinfo{volume}{D90}},
  \bibinfo{pages}{065001} (\bibinfo{year}{2014}{\natexlab{a}}),
  \eprint{1306.6575}.

\bibitem[{\citenamefont{Cachazo
  et~al.}(2014{\natexlab{b}})\citenamefont{Cachazo, He, and
  Yuan}}]{Cachazo:2013hca}
\bibinfo{author}{\bibfnamefont{F.}~\bibnamefont{Cachazo}},
  \bibinfo{author}{\bibfnamefont{S.}~\bibnamefont{He}}, \bibnamefont{and}
  \bibinfo{author}{\bibfnamefont{E.~Y.} \bibnamefont{Yuan}},
  \bibinfo{journal}{Phys. Rev. Lett.} \textbf{\bibinfo{volume}{113}},
  \bibinfo{pages}{171601} (\bibinfo{year}{2014}{\natexlab{b}}),
  \eprint{1307.2199}.

\bibitem[{\citenamefont{Cachazo
  et~al.}(2014{\natexlab{c}})\citenamefont{Cachazo, He, and
  Yuan}}]{Cachazo:2013iea}
\bibinfo{author}{\bibfnamefont{F.}~\bibnamefont{Cachazo}},
  \bibinfo{author}{\bibfnamefont{S.}~\bibnamefont{He}}, \bibnamefont{and}
  \bibinfo{author}{\bibfnamefont{E.~Y.} \bibnamefont{Yuan}},
  \bibinfo{journal}{JHEP} \textbf{\bibinfo{volume}{07}}, \bibinfo{pages}{033}
  (\bibinfo{year}{2014}{\natexlab{c}}), \eprint{1309.0885}.

\bibitem[{\citenamefont{Cachazo
  et~al.}(2015{\natexlab{a}})\citenamefont{Cachazo, He, and
  Yuan}}]{Cachazo:2014xea}
\bibinfo{author}{\bibfnamefont{F.}~\bibnamefont{Cachazo}},
  \bibinfo{author}{\bibfnamefont{S.}~\bibnamefont{He}}, \bibnamefont{and}
  \bibinfo{author}{\bibfnamefont{E.~Y.} \bibnamefont{Yuan}},
  \bibinfo{journal}{JHEP} \textbf{\bibinfo{volume}{07}}, \bibinfo{pages}{149}
  (\bibinfo{year}{2015}{\natexlab{a}}), \eprint{1412.3479}.

\bibitem[{\citenamefont{Arkani-Hamed
  et~al.}(2018{\natexlab{a}})\citenamefont{Arkani-Hamed, Bai, He, and
  Yan}}]{Arkani-Hamed:2017mur}
\bibinfo{author}{\bibfnamefont{N.}~\bibnamefont{Arkani-Hamed}},
  \bibinfo{author}{\bibfnamefont{Y.}~\bibnamefont{Bai}},
  \bibinfo{author}{\bibfnamefont{S.}~\bibnamefont{He}}, \bibnamefont{and}
  \bibinfo{author}{\bibfnamefont{G.}~\bibnamefont{Yan}},
  \bibinfo{journal}{JHEP} \textbf{\bibinfo{volume}{05}}, \bibinfo{pages}{096}
  (\bibinfo{year}{2018}{\natexlab{a}}), \eprint{1711.09102}.

\bibitem[{\citenamefont{Arkani-Hamed et~al.}(2010)\citenamefont{Arkani-Hamed,
  Cachazo, and Kaplan}}]{ArkaniHamed:2008gz}
\bibinfo{author}{\bibfnamefont{N.}~\bibnamefont{Arkani-Hamed}},
  \bibinfo{author}{\bibfnamefont{F.}~\bibnamefont{Cachazo}}, \bibnamefont{and}
  \bibinfo{author}{\bibfnamefont{J.}~\bibnamefont{Kaplan}},
  \bibinfo{journal}{JHEP} \textbf{\bibinfo{volume}{09}}, \bibinfo{pages}{016}
  (\bibinfo{year}{2010}), \eprint{0808.1446}.

\bibitem[{\citenamefont{Cheung et~al.}(2015)\citenamefont{Cheung, Kampf,
  Novotny, and Trnka}}]{Cheung:2014dqa}
\bibinfo{author}{\bibfnamefont{C.}~\bibnamefont{Cheung}},
  \bibinfo{author}{\bibfnamefont{K.}~\bibnamefont{Kampf}},
  \bibinfo{author}{\bibfnamefont{J.}~\bibnamefont{Novotny}}, \bibnamefont{and}
  \bibinfo{author}{\bibfnamefont{J.}~\bibnamefont{Trnka}},
  \bibinfo{journal}{Phys. Rev. Lett.} \textbf{\bibinfo{volume}{114}},
  \bibinfo{pages}{221602} (\bibinfo{year}{2015}), \eprint{1412.4095}.

\bibitem[{\citenamefont{Cachazo
  et~al.}(2015{\natexlab{b}})\citenamefont{Cachazo, He, and
  Yuan}}]{Cachazo:2015ksa}
\bibinfo{author}{\bibfnamefont{F.}~\bibnamefont{Cachazo}},
  \bibinfo{author}{\bibfnamefont{S.}~\bibnamefont{He}}, \bibnamefont{and}
  \bibinfo{author}{\bibfnamefont{E.~Y.} \bibnamefont{Yuan}},
  \bibinfo{journal}{Phys. Rev.} \textbf{\bibinfo{volume}{D92}},
  \bibinfo{pages}{065030} (\bibinfo{year}{2015}{\natexlab{b}}),
  \eprint{1503.04816}.

\bibitem[{\citenamefont{Du and Luo}(2015)}]{Du:2015esa}
\bibinfo{author}{\bibfnamefont{Y.-J.} \bibnamefont{Du}} \bibnamefont{and}
  \bibinfo{author}{\bibfnamefont{H.}~\bibnamefont{Luo}},
  \bibinfo{journal}{JHEP} \textbf{\bibinfo{volume}{08}}, \bibinfo{pages}{058}
  (\bibinfo{year}{2015}), \eprint{1505.04411}.

\bibitem[{\citenamefont{Du and Luo}(2017)}]{Du:2016njc}
\bibinfo{author}{\bibfnamefont{Y.-J.} \bibnamefont{Du}} \bibnamefont{and}
  \bibinfo{author}{\bibfnamefont{H.}~\bibnamefont{Luo}},
  \bibinfo{journal}{JHEP} \textbf{\bibinfo{volume}{03}}, \bibinfo{pages}{062}
  (\bibinfo{year}{2017}), \eprint{1611.07479}.

\bibitem[{\citenamefont{Arkani-Hamed
  et~al.}(2018{\natexlab{b}})\citenamefont{Arkani-Hamed, Rodina, and
  Trnka}}]{Arkani-Hamed:2016rak}
\bibinfo{author}{\bibfnamefont{N.}~\bibnamefont{Arkani-Hamed}},
  \bibinfo{author}{\bibfnamefont{L.}~\bibnamefont{Rodina}}, \bibnamefont{and}
  \bibinfo{author}{\bibfnamefont{J.}~\bibnamefont{Trnka}},
  \bibinfo{journal}{Phys. Rev. Lett.} \textbf{\bibinfo{volume}{120}},
  \bibinfo{pages}{231602} (\bibinfo{year}{2018}{\natexlab{b}}),
  \eprint{1612.02797}.

\bibitem[{\citenamefont{Rodina}(2019{\natexlab{a}})}]{Rodina:2016jyz}
\bibinfo{author}{\bibfnamefont{L.}~\bibnamefont{Rodina}},
  \bibinfo{journal}{JHEP} \textbf{\bibinfo{volume}{09}}, \bibinfo{pages}{084}
  (\bibinfo{year}{2019}{\natexlab{a}}), \eprint{1612.06342}.

\bibitem[{\citenamefont{Strominger}(2017)}]{Strominger:2017zoo}
\bibinfo{author}{\bibfnamefont{A.}~\bibnamefont{Strominger}}
  (\bibinfo{year}{2017}), \eprint{1703.05448}.

\bibitem[{\citenamefont{Zlotnikov}(2017)}]{Zlotnikov:2017ahq}
\bibinfo{author}{\bibfnamefont{M.}~\bibnamefont{Zlotnikov}},
  \bibinfo{journal}{JHEP} \textbf{\bibinfo{volume}{10}}, \bibinfo{pages}{209}
  (\bibinfo{year}{2017}), \eprint{1708.05016}.

\bibitem[{\citenamefont{Rodina}(2019{\natexlab{b}})}]{Rodina:2018pcb}
\bibinfo{author}{\bibfnamefont{L.}~\bibnamefont{Rodina}},
  \bibinfo{journal}{Phys. Rev. Lett.} \textbf{\bibinfo{volume}{122}},
  \bibinfo{pages}{071601} (\bibinfo{year}{2019}{\natexlab{b}}),
  \eprint{1807.09738}.

\bibitem[{\citenamefont{Yin}(2019)}]{Yin:2018hht}
\bibinfo{author}{\bibfnamefont{Z.}~\bibnamefont{Yin}}, \bibinfo{journal}{JHEP}
  \textbf{\bibinfo{volume}{03}}, \bibinfo{pages}{158} (\bibinfo{year}{2019}),
  \eprint{1810.07186}.

\bibitem[{\citenamefont{Carrasco and Rodina}(2019)}]{Carrasco:2019qwr}
\bibinfo{author}{\bibfnamefont{J.~J.~M.} \bibnamefont{Carrasco}}
  \bibnamefont{and} \bibinfo{author}{\bibfnamefont{L.}~\bibnamefont{Rodina}},
  \bibinfo{journal}{Phys. Rev.} \textbf{\bibinfo{volume}{D100}},
  \bibinfo{pages}{125007} (\bibinfo{year}{2019}), \eprint{1908.08033}.

\bibitem[{\citenamefont{Kampf et~al.}(2020)\citenamefont{Kampf, Novotny,
  Shifman, and Trnka}}]{Kampf:2019mcd}
\bibinfo{author}{\bibfnamefont{K.}~\bibnamefont{Kampf}},
  \bibinfo{author}{\bibfnamefont{J.}~\bibnamefont{Novotny}},
  \bibinfo{author}{\bibfnamefont{M.}~\bibnamefont{Shifman}}, \bibnamefont{and}
  \bibinfo{author}{\bibfnamefont{J.}~\bibnamefont{Trnka}},
  \bibinfo{journal}{Phys. Rev. Lett.} \textbf{\bibinfo{volume}{124}},
  \bibinfo{pages}{111601} (\bibinfo{year}{2020}), \eprint{1910.04766}.

\bibitem[{\citenamefont{Elvang and Huang}(2013)}]{Elvang:2013cua}
\bibinfo{author}{\bibfnamefont{H.}~\bibnamefont{Elvang}} \bibnamefont{and}
  \bibinfo{author}{\bibfnamefont{Y.-t.} \bibnamefont{Huang}}
  (\bibinfo{year}{2013}), \eprint{1308.1697}.

\bibitem[{\citenamefont{Cheung et~al.}(2018{\natexlab{b}})\citenamefont{Cheung,
  Shen, and Wen}}]{Cheung:2017ems}
\bibinfo{author}{\bibfnamefont{C.}~\bibnamefont{Cheung}},
  \bibinfo{author}{\bibfnamefont{C.-H.} \bibnamefont{Shen}}, \bibnamefont{and}
  \bibinfo{author}{\bibfnamefont{C.}~\bibnamefont{Wen}},
  \bibinfo{journal}{JHEP} \textbf{\bibinfo{volume}{02}}, \bibinfo{pages}{095}
  (\bibinfo{year}{2018}{\natexlab{b}}), \eprint{1705.03025}.

\bibitem[{\citenamefont{Du and Zhang}(2018)}]{Du:2018khm}
\bibinfo{author}{\bibfnamefont{Y.-J.} \bibnamefont{Du}} \bibnamefont{and}
  \bibinfo{author}{\bibfnamefont{Y.}~\bibnamefont{Zhang}},
  \bibinfo{journal}{JHEP} \textbf{\bibinfo{volume}{07}}, \bibinfo{pages}{177}
  (\bibinfo{year}{2018}), \eprint{1803.01701}.

\bibitem[{\citenamefont{Zhou and Feng}(2018)}]{Zhou:2018wvn}
\bibinfo{author}{\bibfnamefont{K.}~\bibnamefont{Zhou}} \bibnamefont{and}
  \bibinfo{author}{\bibfnamefont{B.}~\bibnamefont{Feng}},
  \bibinfo{journal}{JHEP} \textbf{\bibinfo{volume}{09}}, \bibinfo{pages}{160}
  (\bibinfo{year}{2018}), \eprint{1808.06835}.

\bibitem[{\citenamefont{Bollmann and Ferro}(2019)}]{Bollmann:2018edb}
\bibinfo{author}{\bibfnamefont{M.}~\bibnamefont{Bollmann}} \bibnamefont{and}
  \bibinfo{author}{\bibfnamefont{L.}~\bibnamefont{Ferro}},
  \bibinfo{journal}{JHEP} \textbf{\bibinfo{volume}{01}}, \bibinfo{pages}{180}
  (\bibinfo{year}{2019}), \eprint{1808.07451}.

\bibitem[{\citenamefont{Bjerrum-Bohr et~al.}(2019)\citenamefont{Bjerrum-Bohr,
  Gomez, and Helset}}]{Bjerrum-Bohr:2018jqe}
\bibinfo{author}{\bibfnamefont{N.~E.~J.} \bibnamefont{Bjerrum-Bohr}},
  \bibinfo{author}{\bibfnamefont{H.}~\bibnamefont{Gomez}}, \bibnamefont{and}
  \bibinfo{author}{\bibfnamefont{A.}~\bibnamefont{Helset}},
  \bibinfo{journal}{Phys. Rev.} \textbf{\bibinfo{volume}{D99}},
  \bibinfo{pages}{045009} (\bibinfo{year}{2019}), \eprint{1811.06024}.

\bibitem[{\citenamefont{Gomez and Helset}(2019)}]{Gomez:2019cik}
\bibinfo{author}{\bibfnamefont{H.}~\bibnamefont{Gomez}} \bibnamefont{and}
  \bibinfo{author}{\bibfnamefont{A.}~\bibnamefont{Helset}},
  \bibinfo{journal}{JHEP} \textbf{\bibinfo{volume}{05}}, \bibinfo{pages}{129}
  (\bibinfo{year}{2019}), \eprint{1902.02633}.

\bibitem[{\citenamefont{Zhou}(2019)}]{Zhou:2019mbe}
\bibinfo{author}{\bibfnamefont{K.}~\bibnamefont{Zhou}}, \bibinfo{journal}{JHEP}
  \textbf{\bibinfo{volume}{10}}, \bibinfo{pages}{195} (\bibinfo{year}{2019}),
  \eprint{1908.10272}.

\bibitem[{\citenamefont{Low and Yin}(2019)}]{Low:2019ynd}
\bibinfo{author}{\bibfnamefont{I.}~\bibnamefont{Low}} \bibnamefont{and}
  \bibinfo{author}{\bibfnamefont{Z.}~\bibnamefont{Yin}},
  \bibinfo{journal}{JHEP} \textbf{\bibinfo{volume}{11}}, \bibinfo{pages}{078}
  (\bibinfo{year}{2019}), \eprint{1904.12859}.

\bibitem[{\citenamefont{Carrillo~González
  et~al.}(2019)\citenamefont{Carrillo~González, Penco, and
  Trodden}}]{Carrillo-Gonzalez:2019aao}
\bibinfo{author}{\bibfnamefont{M.}~\bibnamefont{Carrillo~González}},
  \bibinfo{author}{\bibfnamefont{R.}~\bibnamefont{Penco}}, \bibnamefont{and}
  \bibinfo{author}{\bibfnamefont{M.}~\bibnamefont{Trodden}}
  (\bibinfo{year}{2019}), \eprint{1908.07531}.

\bibitem[{\citenamefont{Carrasco
  et~al.}(2017{\natexlab{a}})\citenamefont{Carrasco, Mafra, and
  Schlotterer}}]{Carrasco:2016ldy}
\bibinfo{author}{\bibfnamefont{J.~J.~M.} \bibnamefont{Carrasco}},
  \bibinfo{author}{\bibfnamefont{C.~R.} \bibnamefont{Mafra}}, \bibnamefont{and}
  \bibinfo{author}{\bibfnamefont{O.}~\bibnamefont{Schlotterer}},
  \bibinfo{journal}{JHEP} \textbf{\bibinfo{volume}{06}}, \bibinfo{pages}{093}
  (\bibinfo{year}{2017}{\natexlab{a}}), \eprint{1608.02569}.

\bibitem[{\citenamefont{Carrasco
  et~al.}(2017{\natexlab{b}})\citenamefont{Carrasco, Mafra, and
  Schlotterer}}]{Carrasco:2016ygv}
\bibinfo{author}{\bibfnamefont{J.~J.~M.} \bibnamefont{Carrasco}},
  \bibinfo{author}{\bibfnamefont{C.~R.} \bibnamefont{Mafra}}, \bibnamefont{and}
  \bibinfo{author}{\bibfnamefont{O.}~\bibnamefont{Schlotterer}},
  \bibinfo{journal}{JHEP} \textbf{\bibinfo{volume}{08}}, \bibinfo{pages}{135}
  (\bibinfo{year}{2017}{\natexlab{b}}), \eprint{1612.06446}.

\bibitem[{\citenamefont{Carrasco et~al.}(2019)\citenamefont{Carrasco, Rodina,
  Yin, and Zekioglu}}]{Carrasco:2019yyn}
\bibinfo{author}{\bibfnamefont{J.~J.~M.} \bibnamefont{Carrasco}},
  \bibinfo{author}{\bibfnamefont{L.}~\bibnamefont{Rodina}},
  \bibinfo{author}{\bibfnamefont{Z.}~\bibnamefont{Yin}}, \bibnamefont{and}
  \bibinfo{author}{\bibfnamefont{S.}~\bibnamefont{Zekioglu}}
  (\bibinfo{year}{2019}), \eprint{1910.12850}.

\bibitem[{\citenamefont{Del~Duca et~al.}(2000)\citenamefont{Del~Duca, Dixon,
  and Maltoni}}]{DelDuca:1999rs}
\bibinfo{author}{\bibfnamefont{V.}~\bibnamefont{Del~Duca}},
  \bibinfo{author}{\bibfnamefont{L.~J.} \bibnamefont{Dixon}}, \bibnamefont{and}
  \bibinfo{author}{\bibfnamefont{F.}~\bibnamefont{Maltoni}},
  \bibinfo{journal}{Nucl. Phys.} \textbf{\bibinfo{volume}{B571}},
  \bibinfo{pages}{51} (\bibinfo{year}{2000}), \eprint{hep-ph/9910563}.

\bibitem[{\citenamefont{Kampf et~al.}(2013{\natexlab{a}})\citenamefont{Kampf,
  Novotny, and Trnka}}]{Kampf:2012fn}
\bibinfo{author}{\bibfnamefont{K.}~\bibnamefont{Kampf}},
  \bibinfo{author}{\bibfnamefont{J.}~\bibnamefont{Novotny}}, \bibnamefont{and}
  \bibinfo{author}{\bibfnamefont{J.}~\bibnamefont{Trnka}},
  \bibinfo{journal}{Phys. Rev.} \textbf{\bibinfo{volume}{D87}},
  \bibinfo{pages}{081701} (\bibinfo{year}{2013}{\natexlab{a}}),
  \eprint{1212.5224}.

\bibitem[{\citenamefont{Kampf et~al.}(2013{\natexlab{b}})\citenamefont{Kampf,
  Novotny, and Trnka}}]{Kampf:2013vha}
\bibinfo{author}{\bibfnamefont{K.}~\bibnamefont{Kampf}},
  \bibinfo{author}{\bibfnamefont{J.}~\bibnamefont{Novotny}}, \bibnamefont{and}
  \bibinfo{author}{\bibfnamefont{J.}~\bibnamefont{Trnka}},
  \bibinfo{journal}{JHEP} \textbf{\bibinfo{volume}{05}}, \bibinfo{pages}{032}
  (\bibinfo{year}{2013}{\natexlab{b}}), \eprint{1304.3048}.

\bibitem[{\citenamefont{Bijnens et~al.}(2019)\citenamefont{Bijnens, Kampf, and
  Sjö}}]{Bijnens:2019eze}
\bibinfo{author}{\bibfnamefont{J.}~\bibnamefont{Bijnens}},
  \bibinfo{author}{\bibfnamefont{K.}~\bibnamefont{Kampf}}, \bibnamefont{and}
  \bibinfo{author}{\bibfnamefont{M.}~\bibnamefont{Sjö}},
  \bibinfo{journal}{JHEP} \textbf{\bibinfo{volume}{11}}, \bibinfo{pages}{074}
  (\bibinfo{year}{2019}), \eprint{1909.13684}.

\bibitem[{\citenamefont{Kleiss and Kuijf}(1989)}]{Kleiss:1988ne}
\bibinfo{author}{\bibfnamefont{R.}~\bibnamefont{Kleiss}} \bibnamefont{and}
  \bibinfo{author}{\bibfnamefont{H.}~\bibnamefont{Kuijf}},
  \bibinfo{journal}{Nucl. Phys.} \textbf{\bibinfo{volume}{B312}},
  \bibinfo{pages}{616} (\bibinfo{year}{1989}).

\bibitem[{\citenamefont{Du and Fu}(2016)}]{Du:2016tbc}
\bibinfo{author}{\bibfnamefont{Y.-J.} \bibnamefont{Du}} \bibnamefont{and}
  \bibinfo{author}{\bibfnamefont{C.-H.} \bibnamefont{Fu}},
  \bibinfo{journal}{JHEP} \textbf{\bibinfo{volume}{09}}, \bibinfo{pages}{174}
  (\bibinfo{year}{2016}), \eprint{1606.05846}.

\bibitem[{\citenamefont{Du and Teng}(2017)}]{Du:2017kpo}
\bibinfo{author}{\bibfnamefont{Y.-J.} \bibnamefont{Du}} \bibnamefont{and}
  \bibinfo{author}{\bibfnamefont{F.}~\bibnamefont{Teng}},
  \bibinfo{journal}{JHEP} \textbf{\bibinfo{volume}{04}}, \bibinfo{pages}{033}
  (\bibinfo{year}{2017}), \eprint{1703.05717}.

\bibitem[{\citenamefont{Kawai et~al.}(1986)\citenamefont{Kawai, Lewellen, and
  Tye}}]{Kawai:1985xq}
\bibinfo{author}{\bibfnamefont{H.}~\bibnamefont{Kawai}},
  \bibinfo{author}{\bibfnamefont{D.~C.} \bibnamefont{Lewellen}},
  \bibnamefont{and} \bibinfo{author}{\bibfnamefont{S.~H.~H.}
  \bibnamefont{Tye}}, \bibinfo{journal}{Nucl. Phys.}
  \textbf{\bibinfo{volume}{B269}}, \bibinfo{pages}{1} (\bibinfo{year}{1986}).

\end{thebibliography}

%\begin{thebibliography}{nn}

%\end{thebibliography}

\end{document}